\documentstyle[12pt]{report}
\def\thebibliography#1{\leftline{\it References}\list
  {[\arabic{enumi}]}{\settowidth\labelwidth{[#1]}\leftmargin\labelwidth
    \advance\leftmargin\labelsep
    \usecounter{enumi}}
    \def\newblock{\hskip .11em plus .33em minus .07em}
    \sloppy\clubpenalty4000\widowpenalty4000}

\begin{document}

\rightline{SU--4240--587}
\rightline{UNITU--THEP--22/1994}
\rightline{October 1994}
\rightline{hep-ph/9410321}
\vskip2cm
\centerline{\Large\bf Resolving Ordering Ambiguities in the
Collective}
\vskip0.5cm
\centerline{\Large\bf Quantization by Particle Conjugation
Constraints}
\vskip2cm
\centerline{J.\ Schechter$^{a}$ and H.\ Weigel$^{a,b}$}
\vskip1cm
\centerline{$^a$Department of Physics, Syracuse University}
\centerline{Syracuse, NY 13244--1130, USA}
\vskip0.5cm
\centerline{$^b$Institute for Theoretical Physics,
T\"ubingen University}
\centerline{D-72076 T\"ubingen, Germany}
\vskip 3cm
\baselineskip=20 true pt
\centerline{\bf ABSTRACT}
\vskip .25cm

We formulate the particle conjugation operation and its convenient
realization as $G$--parity in the framework of several chiral
soliton models. The Skyrme model, the Skyrme model with vector mesons
and the chiral quark model are specifically treated. The vector
and axial vector currents are classified according to their
behavior under $G$--parity. In the soliton sector particle conjugation
constrains {\it a priori} ambiguous orderings of operators in the space
of the collective coordinates. In the Skyrme model with vector mesons
and in a local chiral model with an explicit valence quark this
classification scheme provides consistency conditions for the ordering
of the collective operators appearing in the $1/N_C$ corrections to
the nucleon axial charge and the isovector magnetic moment. These
consistency conditions cause the corrections obtained from an ordinary
perturbation expansion to vanish in the context of the collective
quantization of the static soliton configuration. This conclusion
presumably applies to all local effective chiral models.
\vfill\eject

\normalsize
\baselineskip=14 true pt
\stepcounter{chapter}
\leftline{\large\it 1. Introduction}
\bigskip

In this paper we shall discuss some theoretical constraints on the
vector and axial vector current matrix elements of the nucleon
predicted in various kinds of chiral soliton models
(for review articles see \cite{ho86,ho94}). One might, at first,
think that this is a subject which has been completely
exhausted. However, there are, at least, two recent problems:

i) the ``too small $g_A$ problem" \cite{ho86}

ii) the ``proton spin puzzle" \cite{br88}

\noindent
where such considerations appear crucial. These problems involve
current matrix elements which are of subleading order in an appropriate
perturbation approach to the collective Hamiltonian describing the
nucleon. As such, they turn out to involve operator ordering
ambiguities. Symmetries like iso--spin invariance do not seem to be
able to offer guidance on this question. Here we point out that
particle conjugation symmetry can, as it does in ordinary field
theory, provide operator ordering restrictions. For example, charge
conjugation invariance requires the QED interaction to have the ordered
form $\frac{1}{2}eA^\mu\left({\overline \Psi}\gamma_\mu\Psi-
\Psi^T\gamma_\mu^T{\overline \Psi}^T\right)$.

Recently, the very interesting observation has been made
\cite{wa93}--\cite{wa94} that the too small value of the neutron
beta decay constant $g_A$ predicted in many chiral soliton models
might be dramatically improved by including $1/N_C$ corrections.
This seems satisfying both because it agrees with the quark model
result \cite{ka84} in which $g_A$ should behave like $(N_C+2)/3$ so
sizable corrections are expected and because the corrections fit
the pattern expected from unitarity constraints on pion--nucleon
matrix elements \cite{da94} (which imply that the structure of
the leading and next--to leading order terms is identical). In
spite of these pleasing features, we will see that the imposition
of proper ordering constraints requires that these $1/N_C$ corrections
vanish in certain chiral soliton models.

Actually, a different problem with these corrections has already
been noted \cite{al93}. The equation of motion for the pion fields,
which yields the PCAC (Partially Conserved Axial Current) relation,
is derived from the static energy functional of the soliton. This
functional is of the order $N_C$ without any subleading corrections.
Thus this equation of motion cannot account for $1/N_C$ corrections in
the divergence of the axial current. The pion field develops a
Yukawa--type tail whose amplitude is, due to PCAC, proportional to $g_A$.
This allows one to not only compute $g_A$ by direct integration of the
axial current but also from the large distance behavior of the pion
\cite{ad83}. The latter, however, is completely determined from the
classical equation of motion and hence does not contain the $1/N_C$
corrections. Therefore it has been concluded that these corrections
violate PCAC by about 30\% \cite{al93}. This problem was circumvented by
elevating the derivative of the axial current including the corrections
to the equation of motion via PCAC \cite{al93,ho93}. However, this treatment
is not completely satisfactory since it has not proven possible to
derive the so constructed equation of motion from an action principle.


We shall, for simplicity, work in a theory with just the two lightest
quark flavors present. The relevant currents at the quark level are the
vector and axial vector ones of both isoscalar and and isovector type
\begin{eqnarray}
V_\mu^0&=&\frac{1}{2}\left({\overline q}\gamma_\mu q
-q^T\gamma_\mu^T{\overline q}^T\right), \quad
V_\mu^a=\frac{1}{4}\left({\overline q}\gamma_\mu\tau^a q
-q^T\gamma_\mu^T\tau^a{\overline q}^T\right)
\nonumber \\
A_\mu^0&=&\frac{1}{2}\left({\overline q}\gamma_\mu\gamma_5 q
-q^T\gamma_\mu^T\gamma_5^T{\overline q}^T\right), \quad
A_\mu^a=\frac{1}{4}\left({\overline q}\gamma_\mu\gamma_5\tau^a q
-q^T\gamma_\mu^T\gamma_5^T\tau^a{\overline q}^T\right),
\label{quarkcurr}
\end{eqnarray}
where $q$ represents the quark field. Notice that all of these operators
are hermitean. In dealing with the charge conjugation of objects with
definite isospin transformation properties, it is often useful to define
the concept of $G$--parity \cite{ni51}. Technically the $G$--parity
reflection is defined as charge conjugation followed by a
rotation of angle $\pi$ around the $y$--axis in isospace. This
corresponds to a unitary operator ${\cal G}={\rm e}^{i\pi I_2}{\cal C}$.
Since the $G$--parity of the pion is minus one, the $G$--parity
classification provides a tool to determine whether the number of
pions emerging from a decay process is even ($G$--parity$=1$) or
odd ($G$--parity$=-1$). By construction the $G$--parity is a conserved
quantity for interactions which are invariant under charge conjugation
and isospin rotations. For the current operators in (\ref{quarkcurr})
we have
\begin{eqnarray}
{\cal G}\left(V_\mu^0,V_\mu^a,A_\mu^0,A_\mu^a\right){\cal G}^{-1}=
\left(-V_\mu^0,V_\mu^a,A_\mu^0,-A_\mu^a\right).
\label{gparcurr}
\end{eqnarray}

In chiral models the pions are incorporated via the
non--linear realization $U={\rm exp}(i\mbox{\boldmath $\pi$}\cdot
\mbox{\boldmath $\tau$}/f_\pi)$ where $f_\pi=93$MeV denotes the
weak pion decay constant. Since $\mbox{\boldmath $\pi$}
\rightarrow-\mbox{\boldmath $\pi$}$ under the $G$--transformation
we simply have
\begin{eqnarray}
U(\mbox{\boldmath $r$},t)\
{\buildrel{G}\over\longrightarrow}\
U^{\dag}(\mbox{\boldmath $r$},t).
\label{gparu}
\end{eqnarray}
One should bear in mind that in all quark soliton models the quark
fields are functionals of the chiral field so their behavior under
conjugation is determined from that of $U$.

This paper is organized as follows. In section 2 the behavior of
the symmetry currents in the simple Skyrme model under $G$--parity
reflection will be studied. This will be generalized to models with
vector mesons in section 3. In section 4 we will first discuss the
particle conjugation in the framework of quark soliton models and
then apply these results to the $1/N_C$ corrections to $g_A$ and $\mu_V$.

Actually the problem of ordering ambiguities in soliton models
was already discussed by Tomboulis \cite{to75} in the case of a two
dimensional field theory. Also in that reference symmetry arguments
(Poincar\'e invariance) have been employed to resolve operator
ordering ambiguities. However, we will mainly be concerned with form
factors at zero momentum transfer and hence do not introduce
collective coordinates for the boost of the soliton. Rather we are
interested in ordering ambiguities for the collective coordinates
associated with internal symmetries\footnote{For a two dimensional
model operator orderings were considered in ref. \cite{do94} with
regard to internal symmetries.}.

\bigskip
\stepcounter{chapter}
\leftline{\large\it 2. $G$--Parity Constraints on the Currents
in the Skyrme Model}
\bigskip

In order to introduce the relevant background and notation we start
with the simple Skyrme model, defined by the Lagrangian
\cite{sk61,ad83}
\begin{eqnarray}
{\cal L}=\frac{f_\pi^2}{4}
{\rm tr}\left(\partial_\mu U \partial^\mu U^{\dag}\right)+
\frac{1}{32e^2}{\rm tr}\left(
\left[\partial_\mu U, \partial_\nu U^{\dag}\right]
\left[\partial^\mu U, \partial^\nu U^{\dag}\right]\right).
\label{skyrme}
\end{eqnarray}
where $e$ is the only free parameter.

The static soliton configuration of the Skyrme model is given by the
celebrated hedgehog {\it ansatz}
\begin{eqnarray}
U_0(\mbox{\boldmath $r$})={\rm exp}\Big[i
\mbox{\boldmath $\tau$}\cdot
{\hat{\mbox{\boldmath $r$}}}F(r)\Big]
\label{hedge}
\end{eqnarray}
which introduces the chiral angle $F(r)$. Substituting this
{\it ansatz} into the Lagrangian (\ref{skyrme}) yields minus
the classical energy functional $E_{\rm cl}$. The soliton
configuration is then obtained by minimizing this functional.
In the Skyrme model the baryon number current is identified
\cite{ad83} with the topological current which classifies the
mappings of the type (\ref{hedge}) with the boundary condition
$F(r){\buildrel{r\rightarrow\infty}\over\longrightarrow}0$.
The baryon number associated with a given chiral angle, $F(r)$,
is obtained as the spatial integral over the time component of
this current
\begin{eqnarray}
B=\frac{2}{\pi}\int_0^\infty dr F^\prime(r)
{\rm sin}^2F(r)=-\frac{1}{\pi}F(0).
\label{bnumber}
\end{eqnarray}
Integer baryon numbers thus correspond to the boundary conditions
$F(0)=n\pi$ with $n$ being a positive or negative integer. In this
context it should be remarked that in topological soliton models
$E_{\rm cl}$ diverges as one deviates from this boundary condition.
Obviously, one crosses from the baryon to the anti--baryon sector
by reversing the sign of the chiral angle, $F(r)$, since not only
the baryon number but also the baryon number density acquires the
opposite sign.

In order to project the soliton onto states with good spin and isopin
one introduces time dependent collective coordinates parametrizing
the corresponding rotations. Due to the symmetry of the hedgehog
{\it ansatz} the rotations in coordinate and iso--space are equivalent.
Hence one approximates the time dependent solutions by \cite{ad83}
\begin{eqnarray}
U(\mbox{\boldmath $r$},t)=
A(t)U_0(\mbox{\boldmath $r$})A^{\dag}(t).
\label{rothedge}
\end{eqnarray}
The collective coordinates are contained in the SU(2) matrix $A(t)$.
 From the defining relation (\ref{gparu}) we observe that the
$G$--operation corresponds to the reversal of the sign of the
chiral angle $F\rightarrow-F$ while keeping the collective
coordinates, $A(t)$, unaltered. According to (\ref{bnumber})
this indicates that in the soliton sector the $G$--parity
transformation actually corresponds to a particle conjugation.

A more transparent parametrization of the collective coordinates
and their time dependence is given in terms of the
adjoint representation matrix, $D_{ab}$, and the angular
velocity, $\mbox{\boldmath $\Omega$}$,
\begin{eqnarray}
D_{ab}=\frac{1}{2}\ {\rm tr}\left(\tau_aA\tau_bA^{\dag}\right)
\qquad {\rm and} \qquad
A^{\dag}\frac{\partial}{\partial t}A=\frac{i}{2}
\mbox{\boldmath $\tau$}\cdot\mbox{\boldmath $\Omega$}.
\label{angvel}
\end{eqnarray}
The collective Lagrangian associated with the isorotating
hedgehog (\ref{rothedge})
\begin{eqnarray}
L=-E_{\rm cl}+\frac{1}{2}\alpha^2\mbox{\boldmath $\Omega$}^2
\label{colag}
\end{eqnarray}
defines the moment of inertia, $\alpha^2$, which
again is a functional of the chiral angle, $F(r)$\footnote{For
explicit expressions of $E_{\rm cl}$ and $\alpha^2$ in various
Skyrme type models the reader should consult ref. \cite{ho94}
and references therein.}. Here we should stress that the classical
mass as well as the moment of inertia do {\it not}
change when the sign of the chiral angle is reversed, {\it i.e.}
when one goes from the baryon to the anti--baryon. Furthermore
both quantities are of the order $N_C$.

In the process of quantizing the collective coordinates, the spin
operator is identified as the quantity canonical to the
angular velocity $\mbox{\boldmath $\Omega$}$
\begin{eqnarray}
\mbox{\boldmath $J$}=\frac{\partial L}
{\partial \mbox{\boldmath $\Omega$}}=
\alpha^2\mbox{\boldmath $\Omega$}.
\label{defspin}
\end{eqnarray}
For what comes later it important to note that
the spin operators act as {\it right} generators in the
collective space
\begin{eqnarray}
[\mbox{\boldmath $J$},A]=\frac{1}{2}A\mbox{\boldmath $\tau$}
\quad \Longrightarrow \quad
[J_i,D_{ab}]=i\epsilon_{ibc}D_{ac}.
\label{comrules}
\end{eqnarray}
Due to the invariance of the static soliton under combined
rotations in coordinate and iso--space the isospin is
related to the spin via the rotation
\begin{eqnarray}
I_a=-D_{ab}J_b.
\label{defispin}
\end{eqnarray}
This restricts the possible eigenstates of the collective
Hamiltonian
\begin{eqnarray}
H=E_{\rm cl}+\frac{\mbox{\boldmath $J$}^2}{2\alpha^2}
=E_{\rm cl}+\frac{J(J+1)}{2\alpha^2}
\label{coham}
\end{eqnarray}
to those with total isospin $I=J$. Since the isospin operators
correspond to the {\it left} generators one may replace
\begin{eqnarray}
D_{ab}\longrightarrow -\frac{4}{3}I_aJ_b
\label{dreduced}
\end{eqnarray}
in the subspace $I=J=1/2$ as a consequence of the Wigner--Eckart
theorem. We therefore do not need any further specification of
the nucleon or anti nucleon wave--functions in the collective
space.

Various static nucleon properties are obtained from the matrix
elements of the vector and axial vector currents (\ref{quarkcurr}).
These correspond in the Skyrme model to the symmetry currents
which couple linearly to external gauge fields introduced in eqn
(\ref{skyrme}). The currents are then given in terms of the chiral
field and its derivatives. Next the rotating hedgehog configuration
(\ref{rothedge}) is substituted and the angular velocity is eliminated
with the quantization prescription (\ref{defspin}). As the moment of
inertia, $\alpha^2$ is of the order $N_C$ it is obvious that each
such substitution reduces the order of $N_C$ by one unit. Finally the
currents are obtained as combinations of collective operators whose
coefficients are radial functions which depend on the chiral angle,
$F(r)$.  These combinations are displayed schematically in table
\ref{ta_curr}. Note that due to
the static nature of the soliton configuration the time ($\mu=0$)
and spatial ($\mu=i$) components of the currents behave differently.
Actually the form of currents displayed in table \ref{ta_curr} is
the most general one at leading order in $1/N_C$ once proper account
is taken the rotational symmetries in coordinate-- and isospace
as well as parity. Additionally one might have a time component for
the axial singlet current $A_0^0=A_0(r){\hat{\mbox{\boldmath $r$}}}
\cdot\mbox{\boldmath $J$}$. However, $A_0(r)=0$ in Skyrme type
models. We have defined the isospin part of the generator of the
isoscalar currents to be $1/N_C$. The proper normalization of the
electromagnetic charges is guaranteed by $\int d^3r V_1(r)=1$ and
$\int d^3r V_3(r)=\alpha^2$. The explicit forms of the radial
functions $V_1(r),..,A_5(r)$ may readily be extracted from the
review articles \cite{ho86}.

We have also listed in table \ref{ta_curr}, the possible sign change
of the currents when one replaces $F\rightarrow-F$ in the explicit
forms of the radial functions. According to our discussion above
this sign change should represent the $G$--parity of the appropriate
current. In every case, this is seen to agree with (\ref{gparcurr})
in the underlying QCD theory.

\begin{table}
\caption{\label{ta_curr}The structure of the symmetry currents
in chiral soliton models, the power of the leading term in
an $1/N_C$ expansion as well as the behavior of the radial
functions under sign reversal of the chiral angle and
their $G$--parity.}
\vspace{0.5cm}
\centerline{
\begin{tabular}{|c|c c c c|}
\hline
Current & Structure &  ${\cal O}(N_C)$ &
$F\leftrightarrow -F$ & $G$--parity \\
\hline
$V_0^0$ & $V_1(r)$     &  0  & $-$  &$- $ \\
$V_i^0$ & $V_2(r)\epsilon_{ijk}
{\hat {r}}_jJ_k/\alpha^2$
&  -1  & $-$ & $-$ \\
$V_0^a$ & $-V_3(r)D_{ab}J_b/\alpha^2=V_3(r)I_a/\alpha^2$
&  0  & $+$  & $+$ \\
$V_i^a$ & $V_4(r)\epsilon_{ijk}
{\hat {r}}_jD_{ak}$
&  1  & $+$  & $+$ \\
\hline
$A_0^0$ & $0$     &    &   & $+$ \\
$A_i^0$ & $\left(A_1(r)\delta_{ij}+
A_2(r){\hat {r}}_i
{\hat {r}}_j\right)
J_j/\alpha^2$ & -1  &$+$ & $+$ \\
$A_0^a$ & $A_3(r)\epsilon_{ijk}{\hat {r}}_i
\left\{D_{aj},J_k\right\}/\alpha^2$
& 0  & $-$ & $-$ \\
$A_i^a$ & $\left(A_4(r)\delta_{ij}+
A_5(r){\hat {r}}_i
{\hat {r}}_j\right)D_{aj} $
& 1  & $-$ & $-$ \\
\hline
\end{tabular}}
\end{table}
It is important to note that in the transition from the classical objects
$D_{ab}$ and $\mbox{\boldmath $\Omega$}$ to the quantum operators
$D_{ab}$ and $\mbox{\boldmath $J$}$ there can be ordering ambiguities
in cases where the product of two operators is required. Such products
appear for the time components of the isovector--vector current,
$V_0^a$, and the isovector--axial vector current, $A_0^a$. For the
former there is no ambiguity because $[D_{ab},J_b]=0$. For the latter
we must choose a symmetric ordering to preserve the hermitean
nature of the current. Using the substitution (\ref{dreduced}) shows
that matrix elements of $A_0^a$ vanish between nucleon states.
This has the desired consequence that the $G$--parity violating
form factor $G_T(\mbox{\boldmath $q$}^2)$ in the Lorentz covariant
decomposition of the matrix elements
\begin{eqnarray}
\langle N^\prime|A_\mu^a|N\rangle=
{\overline u}(\mbox{\boldmath $p$}^\prime)
\left[G_A(\mbox{\boldmath $q$}^2)\gamma_\mu+
\frac{G_p(\mbox{\boldmath $q$}^2)}{2M}q_\mu+
\frac{G_T(\mbox{\boldmath $q$}^2)}{2M}i\sigma_{\mu\nu}q^\nu\right]
\gamma_5\frac{\tau_a}{2}u(\mbox{\boldmath $p$})\ ,\
q_\mu=p_\mu^\prime-p_\mu
\label{axialff}
\end{eqnarray}
is zero in the Skyrme model ({\it cf.} appendix C of
ref.\cite{me87}). If we had incorrectly chosen a different ordering
for the operators, $A_0^a$ would not have vanished between nucleon
states and a non zero $G_T(\mbox{\boldmath $q$}^2)$ form factor would
have resulted, violating $G$--parity. This provides an illustration
of the connection between proper operator ordering and $G$--parity
invariance.  At this stage we see that a symmetric, or hermitean,
ordering of the collective operator is in accordance with the
$G$--parity symmetry.

Returning to the discussion of table \ref{ta_curr} we again stress
that when going from the baryon to the anti--baryon sector, {\it i.e.}
$F\leftrightarrow-F$, the symmetry currents acquire a sign
according to their $G$--parity quantum numbers. This just reflects
the fact that the number of pions (as measured by the power in
which $F$ appears) is either even or odd for a given current.
Of course, all extensions of the model should maintain this
property. Furthermore, operator ordering ambiguities of a current,
which may occur at higher order in the $1/N_C$ expansion, have to
be resolved in such a way that the phase acquired under
$F\leftrightarrow-F$ is determined by the $G$--parity quantum
number of the current.

This prescription has a fruitful application for the axial--singlet
current, which is especially interesting in the context of the
proton spin puzzle. Not long ago it was proposed that an
axial--singlet current of the form
\begin{eqnarray}
A_\mu^0=ic\ {\rm tr}\left(U\partial_\nu U^{\dag}
U\partial_\mu U^{\dag}U\partial^\nu U^{\dag}\right)=
ic\ {\rm tr}\left(p_\nu p_\mu p^\nu\right),
\label{ryzak}
\end{eqnarray}
(resulting from an additional term in (\ref{skyrme}))
would lead to a non--vanishing matrix element between proton
states \cite{ry89}. The anti--Hermi\-tian pseudovector $p_\mu$
is given by $p_\mu=\partial_\mu\xi\xi^{\dag}+
\xi^{\dag}\partial_\mu\xi$ with $\xi$ denoting a root of the
chiral field, {\it i.e.} $U=\xi^2$. The reversal
$F\leftrightarrow-F$ corresponds to $\xi\leftrightarrow\xi^{\dag}$
and thus $p_\mu\leftrightarrow-p_\mu$. It is then obvious \cite{ja88}
that, unless the current (\ref{ryzak}) vanishes identically, it
contradicts the $G$--parity symmetry, {\it cf.} table
\ref{ta_curr}. Let us see, however, how an improper ordering can
lead to a non--vanishing result. As $p_\mu=p_\mu^a\tau_a$
is a vector in iso--space, (\ref{ryzak}) can be expressed as
$2i\epsilon_{abc}p_\nu^ap_\mu^b p^{\nu c}$ which vanishes as long
as the $p_\mu^a$ are considered to be classical objects. Under the
canonical quantization (\ref{defspin}) the angular velocity,
$\mbox{\boldmath $\Omega$}$, which is contained in $p_0$, is replaced
by $\mbox{\boldmath $J$}/\alpha^2$. Then one is tempted to
replace $\epsilon_{abc}\Omega_b\Omega_c$ by
$\epsilon_{abc}J_bJ_c/(\alpha^2)^2=iJ_a/(\alpha^2)^2$.
This in turn would yield a non--vanishing axial--singlet current
\begin{eqnarray}
A_i^0=\frac{2c}{(\alpha^2)^2}F^\prime(r) {\rm sin}^2F(r)
\hat{\mbox{\boldmath $r$}}\cdot\mbox{\boldmath $J$} \hat r_i=
\frac{4\pi^2c}{(\alpha^2)^2}r^2V_1(r)
\hat{\mbox{\boldmath $r$}}\cdot\mbox{\boldmath $J$} \hat r_i.
\label{ryzak1}
\end{eqnarray}
However, (\ref{ryzak1}) is suspicious on {\it a priori} grounds since
it changes sign on $F\rightarrow-F$, in disagreement with the
requirement that $A_i^0$ have positive $G$--parity. We note that
the hermitean ordering
\begin{eqnarray}
\Omega_a\Omega_b\longrightarrow\frac{1}{2}\left\{
\frac{J_a}{\alpha^2}\frac{J_b}{\alpha^2}+
\frac{J_b}{\alpha^2}\frac{J_a}{\alpha^2}\right\}
\label{symord}
\end{eqnarray}
causes the axial--singlet current to vanish, in agreement with
its $G$--parity transformation property. Note also that the
expression (\ref{ryzak1}) is of order $1/N_C^2$ and thus
not of leading order in the $1/N_C$ expansion.

Thus the studies in the simple Skyrme model indicate that an
hermitean ordering resolves the ambiguities, which occur in the
transition from the classical to the quantum level, in agreement with
the constraints imposed by the $G$--parity symmetry.

\bigskip
\stepcounter{chapter}
\leftline{\large\it 3. Including Vector Mesons}
\bigskip

It is well known that the inclusion of vector mesons improves
many predictions of the simple Skyrme model. The too small value
of $g_A$ is slightly increased  but the problem
remains. Therefore one is naturally curious about possible
${\cal O}(1)$ corrections to the leading ${\cal O}(N_C)$
expression for $A_i^a$ (see table \ref{ta_curr}). In this section,
the $G$--parity of the currents in the model with vector mesons
is studied. It is shown that if an arbitrary operator
ordering were to be allowed then ${\cal O}(1)$ corrections to
both $g_A$ and the isovector magnetic moment, $\mu_V$ would appear
to exist. However, such contributions are seen to be $G$--parity
violating and hence must be eliminated by an appropriate
operator ordering.

In the presence of vector mesons the situation is more involved
not only because the static energy functional, $E_{\rm cl}$ contains
these additional fields but also because field components, which
vanish classically, get excited by the collective rotation
(\ref{rothedge}). At this point we will not present the detailed
structure of the Lagrangian we are considering. Its form has already
been proposed a decade ago \cite{ks84}. Also the static soliton
configuration has been constructed \cite{ja88} and the fields
induced by the collective rotation have been computed \cite{me89}.
These allowed for a reasonable description of static baryon
properties even in the three flavor model \cite{pa92}. Here
we rather wish to discuss the $G$--parity properties of the
currents. Besides the pseudovector $p_\mu$, which is defined
after eqn (\ref{ryzak}), we need the isovector--vector
$v_\mu=\xi^{\dag}\partial_\mu\xi-\partial_\mu\xi\xi^{\dag}$.
Obviously $v_\mu$ is even under $G$--parity. The vector meson
fields are most conveniently parametrized in terms of
\begin{eqnarray}
R_\mu=\omega_\mu+\rho_\mu-\frac{i}{2g}v_\mu.
\label{defr}
\end{eqnarray}
The coupling constant $g$ is determined from the
decay width of the process $\rho\rightarrow\pi\pi$. As the
isoscalar--vector field $\omega$ is odd under $G$--parity
no definite $G$--parity quantum number can be attached to
$R_\mu$. The isovector--vector field $\rho_\mu$ possesses
the same quantum numbers as $v_\mu$. In terms of these
quantities (and the field tensor for the vector meson fields)
the symmetry currents of the model under consideration are
displayed in eqn (2.12) of ref.\cite{pa92}.
There are many terms so we will explicitly display here
just two representative ones for the axial vector current:
\begin{eqnarray}
A_\mu^a&=&-\frac{m_V^2}{g}{\rm tr}\left\{\frac{\tau_a}{2}
\left[\xi R_\mu\xi^{\dag}-\xi^{\dag} R_\mu\xi\right]\right\}
\nonumber \\ &&\hspace{0cm}+
\frac{i\gamma_1}{6}\epsilon_{\mu\nu\rho\sigma}
{\rm tr}\Bigg\{\frac{\tau_a}{2}
\Big[\xi\left( R^\nu p^\rho p^\sigma+
p^\nu p^\rho R^\sigma-p^\nu R^\rho p^\sigma\right)\xi^{\dag}
\nonumber \\ &&\hspace{3cm}+
\xi^{\dag}\left( R^\nu p^\rho p^\sigma+
p^\nu p^\rho R^\sigma-p^\nu R^\rho p^\sigma\right)\xi\Big]
\Bigg\}+\ldots\ .
\label{subset}
\end{eqnarray}
Here $m_V$ denotes the vector meson mass while the coupling
constant $\gamma_1$ can be determined from anomalous decays
like $\omega\rightarrow\pi\pi\pi$ \cite{ja88}. Since $g_A$ is
given by the matrix element of $\int d^3r\ 2A_3^3$ between
proton states with spin projection $+\frac{1}{2}$ we only need
to consider the spatial components $A_i^a$. In order to discuss
the $G$--parity properties of $A_i^a$, we need both the static as
well as the excited vector meson fields. The static
hedgehog {\it ans\"atze} read
\begin{eqnarray}
\omega_0(\mbox{\boldmath $r$})=\frac{1}{2g}\omega(r)
\qquad {\rm and} \qquad
\rho_i^a(\mbox{\boldmath $r$})=
\frac{G(r)}{gr}\epsilon_{ika}\hat{r}_k
\label{vmhedge}
\end{eqnarray}
while all other field components vanish. The equations of
motion for the radial functions $\omega(r)$ and $G(r)$ as
obtained from minimizing the classical energy functional
are, in general, non--linear inhomogeneous differential
equations. The source for $\omega(r)$ is of the form
$V_1(r)$ ({\it cf.} table \ref{ta_curr}) while
$G(r)\sim \epsilon_{ika}\hat{r}_k\ {\rm tr}(\tau_av_i)$.
Hence $\omega(r)$ changes its sign under the reversal
$F\rightarrow-F$ but $G(r)$ does not. Thus we find that
again the transformation $F\rightarrow-F$ gives the proper
$G$--parity quantum number. Upon the collective isorotation
(\ref{rothedge}) we have not only $\rho_i^a(\mbox{\boldmath $r$})
\rightarrow A(t)\rho_i^a(\mbox{\boldmath $r$})A^{\dag}(t)$, but, as
already mentioned, additional fields are induced
\begin{eqnarray}
\omega_i=\frac{\Phi(r)}{2g}\epsilon_{ijk}\Omega_j\hat{r}_k
\qquad {\rm and} \qquad
\rho_0=\frac{1}{2g}A(t)\mbox{\boldmath $\tau$}\cdot
\left[\xi_1(r)\mbox{\boldmath $\Omega$}+
\xi_2(r)(\hat{\mbox{\boldmath $r$}}
\cdot\mbox{\boldmath $\Omega$})
\hat{\mbox{\boldmath $r$}}\right])A^{\dag}(t).
\label{vminduced}
\end{eqnarray}
It is important to remark that the ordering between $A(t)$ and
$\mbox{\boldmath $\Omega$}\ (\rightarrow
\mbox{\boldmath $J$}/\alpha^2)$ is completely arbitrary because
they are considered to be commuting $c$--number quantities.
The radial functions $\xi_1(r),\ \xi_2(r)$ and $\Phi(r)$ solve
linear inhomogeneous differential equations which serve to extremize
the moment of inertia, $\alpha^2$. The classical fields act
as inhomogeneous parts in these equations. Hence one can deduce
the behavior of the induced fields under $F\rightarrow-F$. It
turns out that $\xi_1(r)$ and $\xi_2(r)$ are even while
$\Phi(r)$ is odd, in agreement with $G$--parity of these
fields. These results are summarized in table \ref{ta_vm}.
\begin{table}
\caption{\label{ta_vm}The behavior of the vector meson
radial functions under $F\rightarrow-F$ and the $G$--parity
quantum number of these fields.}
\vspace{0.5cm}
\centerline{
\begin{tabular}{|c|c c |c c c|}
\hline
& $\omega(r)$ & $G(r)$ & $\Phi(r)$ & $\xi_1(r)$ & $\xi_2(r)$ \\
\hline
$F\rightarrow-F$ & - & + & - & + & + \\
$G$--parity &- & + & - & + & + \\
\hline
\end{tabular}}
\end{table}
It is, of course, not surprising that the behavior of the
vector mesons under $F\rightarrow-F$ coincides with their
$G$--parity as the underlying Lagrangian possesses this symmetry
and the vector mesons couple according to their isospin and
spin quantum numbers to the currents displayed in table
\ref{ta_curr}.

Now we are completely equipped to discuss possible ordering
ambiguities for $g_A$ and their eventual resolution by demanding
the proper $G$--parity behavior. We will see that up to next--to
leading order in the $1/N_C$ expansion no contradiction to
the $G$--parity symmetry appears as long as we consider all
fields to be classical objects. Due to its isoscalar character
the $\omega$ meson does not contribute in the first term in
eqn (\ref{subset}), which is ``non--anomalous" ({\it i.e.}
no $\epsilon$--symbol). It is thus obvious that this term is
odd under $G$--parity just as the isovector--axial vector current
is supposed to be. For the anomalous term the situation is somewhat
more involved. One can convince oneself that for this term the proper
$G$--parity is obtained only when $R$ is odd, {\it i. e.} only the
$\omega$ meson is allowed. Since we are taking $\mu=i$,
one of the three remaining indices has to be associated with the
time coordinate. Let us first assume that this is attached to the
$\omega$ field, which is classical. The $p$'s are classical as well
because only their spatial components contribute. Thus the total
contribution from the anomalous terms to $A_i^a$ is classical and
no ordering ambiguities appear when the time like coordinate is
assigned to the $\omega$ meson. This, of course, is nothing but the
leading order part of $A_i^a$ in the $1/N_C$ expansion. It
contributes to the radial functions $A_4(r)$ and $A_5(r)$ in
table \ref{ta_curr} and has been studied in detail \cite{me89,pa92}.
What happens when the $\omega$ field is spatial? In that case
the contribution of the anomalous term to $A_i^a$ contains
two powers of $\mbox{\boldmath $\Omega$}$ since one of the
two $p$'s contains a time derivative. Hence this term is of
next--to--next--to leading order in the $1/N_C$ expansion and
is commonly neglected.

A next--to leading order term can be obtained only when the
isovector part of $R^\nu$ is considered and the angular velocity is
treated as an operator in the collective space. Let us see how this
comes about. A straightforward calculation assuming the apparently
natural ordering indicated in (\ref{vminduced}) leads to
\begin{eqnarray}
g_A^{(1)}=\frac{16\pi}{9g}\ \gamma_1 \int dr
\left[rF^\prime\left(G+\xi_1\right){\rm sin}F
-\left(\xi_1+\xi_2\right){\rm sin}^2F\right]
\Biggl<
{\rm tr}\left(\frac{\tau_3}{2}A\Omega_3A^{\dag}\right)
\Biggr>_N
\label{dgavm}
\end{eqnarray}
where the matrix element with respect to the nucleon state
is indicated. Classically the trace in (\ref{dgavm}) vanishes
identically. However, if one replaced $\Omega_3$ by
$J_3/\alpha^2$ this would no longer by the case. Instead one
would find by using (\ref{comrules}) and (\ref{dreduced})
\begin{eqnarray}
\Biggl<
{\rm tr}\left(\frac{\tau_3}{2}A\Omega_3A^{\dag}\right)
\Biggr>_N\rightarrow
\frac{1}{2\alpha^2}\Biggl<
{\rm tr}\left(\frac{\tau_3}{2}\left[A,J_3\right]A^{\dag}\right)
\Biggr>_N=\frac{-1}{4\alpha^2}
\Biggl<D_{33}\Biggr>_N=\frac{-1}{12\alpha^2}.
\label{vmcom}
\end{eqnarray}
However, this result cannot be taken seriously since the integrand
in eqn (\ref{dgavm}) has the incorrect transformation property when
going from the nucleon to the anti--nucleon. According to table
\ref{ta_curr} the integrand should be odd under $F\rightarrow-F$ but
it is even as can be observed from table \ref{ta_vm}. Physically
$g_A^{(1)}\ne0$ implies that the decay amplitude for the $\beta$-decay
of the anti--neutron is different in magnitude from that of the
neutron. This, of course, contradicts our present understanding of
physics. Hence we conclude that the collective operators have
to be ordered such that (\ref{dgavm}) vanishes identically.

Here we have just considered two terms of the complete axial vector
current in the vector meson model. The above arguments, however,
apply as well to all terms in (2.12) of ref.\cite{pa92}.
Furthermore one may conduct the same studies for the
isovector--vector current in this model. The spatial components
of this current give the isovector part of the magnetic moment
$\mu_V$. As expected, the leading order in the $1/N_C$ expansion is
free of ambiguities and agrees with the $G$--parity constraints. At
the next--to leading order ambiguities for the collective operators
occur. Again the improper behavior of the radial part of the
matrix element under $F\rightarrow-F$ forces one to arrange these
operators so that the $1/N_C$ correction to $\mu_V$ vanishes.

The lesson we learn from studying the symmetry currents in the
vector meson model is that matrix elements which are different from
zero only because of the commutation relations (\ref{comrules})
but vanish classically are likely to violate the $G$--parity
symmetry. In that event those matrix elements should be
discarded.

\bigskip
\stepcounter{chapter}
\leftline{\large\it 4. Chiral Quark Solitons}
\bigskip

The results found in the vector meson model make us suspicious about
the recently discovered $1/N_C$ corrections to $g_A$ and $\mu_V$
\cite{wa93}--\cite{wa94} since they depend on operator ordering.
Especially, it has been noticed that both the vector meson addition
and the quark addition to the non--linear $\sigma$ model similarly
describe short distance corrections to the pion cloud. This is true
both for the problems of the neutron--proton mass difference \cite{ja89}
and the matrix element of the axial singlet current \cite{jo90}.

Unfortunately the effect of the reversal $F\rightarrow-F$ is not as
clear in quark soliton models as it is in purely mesonic models. We
therefore divide the present section into two parts. In the first one
we set up the chiral quark model for both the nucleon and the
anti--nucleon sectors. We derive a prescription to transform the quark
spinors between these two sectors ({\it cf.} (\ref{statrev})). In the
second part this is applied to the investigation of the behavior of
the $1/N_C$ corrections to $g_A$ and $\mu_V$ under $F\rightarrow-F$.
Technical details of these studies are presented in appendix A.

The chiral quark model \cite{kr84,jjs88}, the simplest chirally
symmetric model to contain quark solitons, is defined as the sum of
the non--linear $\sigma$--model Lagrangian with a pion mass term
\begin{eqnarray}
{\cal L}_{{\rm nl}\sigma}=\frac{f_\pi^2}{4}{\rm tr}
\left(\partial_\mu U\partial^\mu U^{\dag}\right)
+\frac{m_\pi^2f_\pi^2}{4}{\rm tr}\left(U+U^{\dag}-2\right)
\label{lnls}
\end{eqnarray}
and a valence quark field $\Psi_{\rm val}$ in the background of
the chiral field $U$
\begin{eqnarray}
{\cal L}_{\rm q}= {\overline \Psi}_{\rm val}
\left(i{\partial \hskip -0.5em /}
-mU^{\gamma_5}\right)\Psi_{\rm val}.
\label{lq}
\end{eqnarray}
Here the mass $m=g_{\rm q}f_\pi$ represents a convenient
parametrization of the coupling $g_{\rm q}$ between the valence
quark and the chiral field. In this model the eventual effects
of sea quarks are assumed to be represented by the kinetic term
in (\ref{lnls}). It is also obvious that the model is formulated
locally.

Again we employ the hedgehog {\it ansatz} (\ref{hedge}) for the
pseudoscalar fields. For this static configuration the Dirac equation
becomes an eigenvalue problem
\begin{eqnarray}
h(F)\Psi_{\rm val}=\epsilon_{\rm val}\Psi_{\rm val}
\label{direqn}
\end{eqnarray}
which defines the Dirac Hamiltonian
\begin{eqnarray}
h(F)={\mbox {\boldmath $\alpha$}} \cdot {\mbox{\boldmath $p$}} +
\beta m \Big[{\rm cos}F(r) + i\gamma_5{\mbox{\boldmath $\tau$}}
\cdot{\hat{\mbox{\boldmath $r$}}}\ {\rm sin}F(r)\Big]
\label{hstat}
\end{eqnarray}
as a function of the chiral angle, $F(r)$. This Dirac Hamiltonian
commutes with the grand spin
\begin{eqnarray}
\mbox{\boldmath $G$}=\mbox{\boldmath $l$}+
\frac{\mbox{\boldmath $\tau$}}{2}+
\frac{\mbox{\boldmath $\sigma$}}{2}
\label{gspin}
\end{eqnarray}
and the parity ($\Pi$) operators. Thus the eigenstates are
classified by $G^{\Pi_{\rm int}}$ where $\Pi_{\rm int}$ refers to the
intrinsic parity, which is defined via $\Pi=\Pi_{\rm int}(-1)^G$.
The construction of these eigenstates is described in appendix A.

The eigenvalue $\epsilon_{\rm val}$ turns out to be a functional
of $F$, as does the total classical energy
\begin{eqnarray}
E_{\rm cl}=E_{{\rm nl}\sigma}+{\rm sgn}(B)N_C\epsilon_{\rm val}
\label{ecl}
\end{eqnarray}
with $E_{{\rm nl}\sigma}=2\pi f_\pi^2 \int dr
\left(r^2F^{\prime2}+2{\rm sin}^2F+2m_\pi^2(1-{\rm cos}F)\right)$.
The sign of the baryon number has been
included in order to accommodate the hole interpretation of the
Dirac theory. In a moment we will see that this relative sign is
also obtained by requiring that the nucleon and anti--nucleon possess
equal masses.

The soliton is computed by solving the Euler--Lagrange equation
for the chiral angle
\begin{eqnarray}
F^{\prime\prime}=-\frac{2}{r}F^\prime+
\frac{{\rm sin}2F}{r^2}+m_\pi^2{\rm sin}F
-{\rm sgn}(B)\frac{N_C m}{f_\pi^2}\int \frac{d\Omega}{4\pi}
\Psi_{\rm val}^{\dag}\beta\left({\rm sin}F
-i\gamma_5\hat{\mbox{\boldmath $r$}}\cdot
\mbox{\boldmath $\tau$}{\rm cos}F\right)\Psi_{\rm val}
\label{eleqn}
\end{eqnarray}
self--consistently. For the boundary condition $F(0)=-\pi$ the bound
valence quark state of the self--consistent solution is found in the
$G^{\pi_{\rm int}}=0^+$ channel. We denote the corresponding eigenvalue
and eigenfunction by $\epsilon_{\rm val}^+$ and $\Psi_{\rm val}^+$,
respectively. On the other hand, when assuming the boundary condition
$F(0)=\pi$, a strongly bound quark is only obtained in the $0^-$ channel
with eigenvalue $\epsilon_{\rm val}^-$ and eigenfunction
$\Psi_{\rm val}^-$. This, of course, reflects nothing but the fact
that the parities of the nucleon and the anti--nucleon are opposite. In
fact the self--consistent solutions for the nucleon ($F(0)=-\pi$) and
anti--nucleon ($F(0)=\pi$) are distinguished by different overall signs
of $F(r)$ and the eigenvalues of the bound quarks, {\it i.e.}
$\epsilon_{\rm val}^-=-\epsilon_{\rm val}^+$. Thus, due to the
inclusion of the factor ${\rm sgn}(B)$ in (\ref{ecl}),
the classical masses of the baryon and anti--baryon are identical.

Actually, the self--consistent solution for the anti--baryon can
easily be obtained from the one for the baryon by noting that
\begin{eqnarray}
h(-F)=-{\cal J}^{\dag}h(F){\cal J}
\qquad {\rm with}\qquad
{\cal J}={\cal J}^{\dag}=i\beta\gamma_5.
\label{hrevf}
\end{eqnarray}
The transformation ${\cal J}$ commutes with $\mbox{\boldmath $G$}$ but
has negative parity. This implies for the eigenvalues and eigenstates
of the Dirac Hamiltonian
\begin{eqnarray}
\epsilon_\mu^{G^{\pm}}\ {\buildrel{F\rightarrow -F}
\over\longrightarrow}\ -\epsilon_\mu^{G^{\mp}}
\qquad{\rm and}\qquad
|\mu,G^{\pm}\rangle\
&{\buildrel{F\rightarrow -F}\over\longrightarrow}&\
|\mu,G^{\mp}\rangle={\cal J}|\mu,G^{\pm}\rangle,
\label{statrev}
\end{eqnarray}
where $\mu$ labels the particular eigenstate. Details of the
transformation (\ref{statrev}) in terms of the radial parts of the
quark wave--functions are given in appendix A. We will make
extensive use of this transformation when comparing the currents
for the nucleon and anti--nucleon. However, we first have to
perform the projection of the chiral quark soliton onto states with
good spin and isospin. For the non--linear $\sigma$ model this is
straightforwardly achieved by substituting the rotating hedgehog
(\ref{rothedge}) into the defining equation (\ref{lnls}). This yields
the mesonic part of the moment of inertia
\begin{eqnarray}
\alpha^2_{\rm m}=\frac{8\pi}{3}f_\pi^2\int dr r^2 {\rm sin}^2F
\label{almes}
\end{eqnarray}
which obviously is identical for the baryon and the anti--baryon. In
order to compute the quarks' contribution to $\alpha^2$ we employ the
cranking method \cite{in54}. Then the quark spinors are rotating in
isospace, which adds the Coriolis term
$(1/2)\mbox{\boldmath $\tau$}\cdot\mbox{\boldmath $\Omega$}$ as a
perturbation to the Dirac Hamiltonian (\ref{hstat}). The energy
eigenvalue acquires a change in second order perturbation, which (due
to isospin invariance) may be written as
$\epsilon_{\rm val}\rightarrow\epsilon_{\rm val}
+(1/2)\alpha^2_{\rm q}\mbox{\boldmath $\Omega$}^2$. Here
\begin{eqnarray}
\alpha^2_{\rm q}=\alpha^2_{\rm q}[F]=\frac{N_C}{2}
\sum_{\mu\ne{\rm val}}
\frac{\left|\langle\mu|\tau_3|{\rm val}\rangle\right|^2}
{\epsilon_{\rm val}^+-\epsilon_\mu}
\label{alquark}
\end{eqnarray}
denotes the quarks' contribution to the moment of inertia of the
nucleon. As $[{\cal J},\tau_i]=0$ one finds that $\alpha^2_{\rm q}[F]$
changes sign under the transformation (\ref{statrev}),
$\alpha^2_{\rm q}[-F]=-\alpha^2_{\rm q}[F]$. Taking into account
that (\ref{alquark}) represents a perturbation to valence quark
energy, which enters the total energy with the overall factor
${\rm sgn}(B)$, the total moment of inertia
\begin{eqnarray}
\alpha^2=\alpha^2_{\rm m}+{\rm sgn}(B)\alpha^2_{\rm q}
\label{altot}
\end{eqnarray}
is seen to be invariant under $F\rightarrow-F$. This guarantees
equal masses for the baryons and anti--baryons in the presence of
rotational corrections.

In the context of the cranking method the induced components of
the valence quarks are computed in a perturbation expansion in
powers of $\mbox{\boldmath $\Omega$}$. This will serve to make
the discussion of the particle conjugation properties
of their contribution to the currents more
transparent than in a variational approach \cite{co86}. In order
to be consistent with the expansion for the energy we require the
first order expression for the cranked wave--function
\begin{eqnarray}
\Psi_{\rm crank}=A(t)\left\{\Psi_{\rm val}
+\frac{1}{2}\sum_{\mu\ne{\rm val}}\Psi_\mu
\frac{\langle\mu|\mbox{\boldmath $\tau$}\cdot
\mbox{\boldmath $\Omega$}|{\rm val}\rangle}
{\epsilon_{\rm val}-\epsilon_\mu}\right\}.
\label{delval}
\end{eqnarray}
The isospin matrices $\mbox{\boldmath $\tau$}$ carry unit grand
spin and are positive under parity. Thus $|\mu\rangle\in 1^\pm$
for $F(0)=\pm\pi$.

We now turn to the discussion of the currents in the chiral
quark model. These are again sums of mesonic and quark parts. As they
are the Noether currents of the underlying theory the overall factor
${\rm sgn}(B)$ is carried along for the quark part
\begin{eqnarray}
V_\mu^a(A_\mu^a)=\frac{i}{2}f_\pi^2{\rm tr}
\left\{\frac{\tau_a}{2}\left(\xi p_\mu \xi^{\dag} \mp
\xi^{\dag} p_\mu \xi\right)\right\}
+{\rm sgn}(B){\overline \Psi}_{\rm crank}\gamma_\mu(\gamma_5)
\frac{\tau^a}{2}\Psi_{\rm crank}.
\label{currclq}
\end{eqnarray}
The interesting part is, of course, the one which is due to
the quarks, especially the spatial components (${\cal
Q}_i^a=\alpha_i(\sigma_i)\ \tau_a/2$)
\begin{eqnarray}
N_C\Psi^{\dag}_{\rm crank}{\cal Q}_i^b\Psi_{\rm crank}&=&
N_C D_{ab}\Psi^{\dag}_{\rm val}{\cal Q}_i^b\Psi_{\rm val}
\label{delcur} \\
&&\hspace{-2cm}
+\frac{N_C}{2}(D_{ab}\Omega_j)\sum_{\mu\ne{\rm val}}
\frac{\langle\mu|\tau_j|{\rm val}\rangle}
{\epsilon_{\rm val}-\epsilon_\mu}
\Psi^{\dag}_{\rm val}{\cal Q}_i^b\Psi_\mu
+\frac{N_C}{2}(\Omega_jD_{ab})\sum_{\mu\ne{\rm val}}
\frac{\langle{\rm val}|\tau_j|\mu\rangle}
{\epsilon_{\rm val}-\epsilon_\mu}
\Psi^{\dag}_\mu{\cal Q}_i^b\Psi_{\rm val}.
\nonumber
\end{eqnarray}
In this expression we have temporarily adopted the ordering
between $D_{ab}$ and $\mbox{\boldmath $\Omega$}$ suggested by the
form of the perturbed wave--function (\ref{delval}). In the framework
of the chiral quark model the $1/N_C$ corrections to $g_A$ and $\mu_V$,
which were recently discussed in various similar models
\cite{wa93,bl93,ho93,ch94,wa94}, correspond to the replacement
$\mbox{\boldmath $\Omega$}\rightarrow\mbox{\boldmath $J$}/\alpha^2$
in the ordering of (\ref{delcur}).

To obtain $g_A$ we calculate the integral
$\int d^3r A_3^3$. The contribution from the next to leading order
involves the sum over the grand spin projection (in which the
energy eigenvalues are degenerate)
\begin{eqnarray}
\sum_M\langle{\rm val,\pm}|\sigma_3\tau_b|\mu,1M,\mp\rangle
\langle\mu,1M,\mp|\tau_j|{\rm val,\pm}\rangle&=&
\frac{i}{3}S_\mu^\pm L_\mu^\pm\epsilon_{3bj}.
\label{sproj}
\end{eqnarray}
The signs appearing in the state vectors label the intrinsic parity.
The upper (lower) sign refers to the (anti) nucleon sector.
We refer to appendix A for the actual computation of these matrix
elements and the definition of the integrals on the $RHS$. Since these
matrix elements are proportional to the anti--symmetric
$\epsilon$--tensor the appearance of the commutator $[D_{ab},J_j]$
is obvious (see (\ref{comrules})) when replacing the angular
velocity by the spin operator. Taking account of this prescription one
finds for the axial charge of either the nucleon or the anti--nucleon
\begin{eqnarray}
g_A=g_A^{(0)}+g_A^{(1)}&=&\frac{8\pi}{9}f_\pi^2\int dr r^2\
\left(F^\prime+\frac{{\rm sin}F}{r}\right)
\label{gachqu} \\ &&
-{\rm sgn}(B)\frac{N_C}{3}\left(\langle{\rm val}|
\sigma_3\tau_3|{\rm val}\rangle+\frac{i}{\alpha^2}
\sum_{\mu\ne{\rm val}}\frac{\langle{\rm val}|\tau_1|\mu\rangle
\langle\mu|\sigma_3\tau_2|{\rm val}\rangle}
{\epsilon_{\rm val}-\epsilon_\mu}\right).
\nonumber
\end{eqnarray}
where the $1/N_C$ correction, $g_A^{(1)}$, is represented by the term
proportional to $1/\alpha^2$. We should remark that (\ref{gachqu})
corresponds to a spin up proton or a spin up anti--neutron which each
give the same matrix element of $D_{33}$. Note that the overall sign
differs in the two cases. Similar calculations give for the isovector
part of the magnetic moment
\begin{eqnarray}
\mu_V=\mu_V^{(0)}+\mu_V^{(1)}&=&\frac{8\pi}{9}f_\pi^2M_N
\int dr r^3 {\rm sin}^2F
\label{mvchqu} \\ && \hspace{-3cm}
-{\rm sgn}(B)\frac{iN_C}{12}M_N\left(\langle{\rm val}|
\left[\mbox{\boldmath $r$}\cdot\mbox{\boldmath $\alpha$},
\sigma_3\right]\tau_3|{\rm val}\rangle+\frac{i}{\alpha^2}
\sum_{\mu\ne{\rm val}}
\frac{\langle{\rm val}|\tau_1|\mu\rangle
\langle\mu|\left[\mbox{\boldmath $r$}\cdot\mbox{\boldmath $\alpha$},
\sigma_3\right]\tau_2|{\rm val}\rangle}
{\epsilon_{\rm val}-\epsilon_\mu}\right).
\nonumber
\end{eqnarray}
Here the $\epsilon$--tensor, which again causes the appearance of the
commutator (\ref{comrules}) in the next--to leading term, originates
from the definition of the isovector part of the magnetic moment
$\mu_V=1/2\langle(\mbox{\boldmath $r$}\times
\mbox{\boldmath $V$}^3\rangle$.

We have computed these matrix elements in the chiral quark model
numerically. It turns out the next--to leading order is only about
$1/10$ of the leading order. Thus these terms are more strongly
suppressed than in the NJL model. This is apparently due to the fact
that the chiral angle in the chiral quark model has quite a large
extension yielding a larger moment of inertia.

Now we can investigate these expressions with respect to their
behavior under particle conjugation. This is made simple by noting
that we just have to apply the transformation (\ref{statrev}). Since
${\cal J}\alpha_i=-\alpha_i{\cal J}$ and
${\cal J}\sigma_i=\sigma_i{\cal J}$ it is obvious (noting the factor
${\rm sgn}(B)$) that the leading order terms transform properly in
agreement with table \ref{ta_curr},
{\it i.e}
\begin{eqnarray}
\langle{\rm val},+|\sigma_3\tau_3|{\rm val},+\rangle
&{\buildrel{F\rightarrow -F}\over\longrightarrow}&\
\langle{\rm val},-|\sigma_3\tau_3|{\rm val},-\rangle,
\label{leading}\\
\langle{\rm val},+|\left[\mbox{\boldmath $r$}\cdot
\mbox{\boldmath $\alpha$},\sigma_3\right]\tau_3
|{\rm val},+\rangle
&{\buildrel{F\rightarrow -F}\over\longrightarrow}&\
-\langle{\rm val},-|\left[\mbox{\boldmath $r$}\cdot
\mbox{\boldmath $\alpha$},\sigma_3\right]\tau_3
|{\rm val},-\rangle,
\end{eqnarray}
Since the transformation ${\cal J}$ does not affect the
isospin structure of the matrix elements it is obvious
that the matrix elements involved in the terms describing the
$1/N_C$ corrections transform in the same way
\begin{eqnarray}
\langle\mu,-|\tau_1|{\rm val},+\rangle\hspace{-0.5pt}
\langle{\rm val,+}|\sigma_3\tau_2|\mu,-\rangle
&{\buildrel{F\rightarrow -F}\over\longrightarrow}&\
\langle\mu,+|\tau_1|{\rm val,-}\rangle\hspace{-0.5pt}
\langle{\rm val,-}|\sigma_3\tau_2|\mu,+\rangle
\label{next} \\
\langle\mu,+|\tau_1|{\rm val},-\rangle\hspace{-0.5pt}
\langle{\rm val},+|\left[\mbox{\boldmath $r$}\cdot
\mbox{\boldmath $\alpha$},\sigma_3\right]\tau_2|\mu,-\rangle
&{\buildrel{F\rightarrow-F}\over\longrightarrow}&
-\langle\mu,+|\tau_1|{\rm val},-\rangle\hspace{-0.5pt}
\langle{\rm val},-|\left[\mbox{\boldmath $r$}\cdot
\mbox{\boldmath $\alpha$},\sigma_3\right]\tau_2|\mu,+\rangle.
\nonumber
\end{eqnarray}
In appendix B we present an additional argument, which is
based on the completeness of the eigenstates of the Dirac
Hamiltonian, for the transformation relation (\ref{next}).
Now the $G$--parity violation of the $1/N_C$ corrections
becomes clear: The energy denominator in these terms
changes its sign when going from the nucleon to the
anti--nucleon. Thus the $1/N_C$ corrections transform
oppositely to the leading order contribution, {\it i.e.}
incorrectly. We therefore conclude that the ordering
ambiguities contained in the formulation of the quark
currents (\ref{delcur}) have to be resolved in such a way that
the $1/N_C$ corrections to $g_A$ and $\mu_V$ vanish:
\begin{eqnarray}
g_A^{(1)}=0
\qquad {\rm and}\qquad
\mu_V^{(1)}=0.
\label{finconcl}
\end{eqnarray}
This can be achieved by demanding an hermitean ordering
prescription
\begin{eqnarray}
D_{ab}\Omega_j\longrightarrow
\frac{1}{2\alpha^2}\left(D_{ab} J_j +
J_j D_{ab}\right).
\label{hermord}
\end{eqnarray}
For this ordering each of the two subleading terms in
(\ref{currclq}) vanishes in the nucleon subspace after
performing the sum over the grand spin projection (\ref{sproj}).
Since
\begin{eqnarray}
\epsilon_{kbj}\left(D_{ab}J_j+J_jD_{ab}\right)
=-\frac{4}{3}I_a\epsilon_{kbj}\left(J_b J_j+J_j J_b\right)=0
\label{final}
\end{eqnarray}
the required result (\ref{finconcl}) is obtained.

It is seen that the pattern obtained in the vector meson model is
repeated in the chiral quark model. As already noted we are considering
a local formulation of the chiral quark model. Non--local effects, as
{\it e.g.} the blocking of transitions to negative energy states in
(\ref{delval}), associated with the inclusion of levels other than
$|{\rm val}\rangle$ in the definition (\ref{lq}) may cause different
expressions for $g_A^{(1)}$ and $\mu_V^{(1)}$. We will comment on
related studies \cite{bl93,ch94} below.

To avoid confusion, we should remark that there is no
{\it practical} problem in the chiral quark model
with too small $g_A$. In fact, $g_A$ comes out slightly too large
in this model, although the problem can be resolved \cite{jjs88}by
complicating the model a bit. However, the structure of the chiral
quark model is, for present purposes, the same as other quark soliton
models.

Let us conclude this section by commenting on the $G$--parity
behavior of the axial singlet current in the chiral quark model.
This has been noticed some time ago to account for the proton
spin puzzle fairly well \cite{jo90}. When the isospin part
of the generator ${\cal Q}_i^a$ reduces to a unit matrix, the
first term on the $RHS$ of (\ref{delcur}) is easily observed
to vanish. Furthermore no ordering ambiguities occur in the
second and third terms because the $D$--matrix is replaced by
the unit matrix. The radial part of the matrix element between
nucleon states of the axial singlet current
\begin{eqnarray}
\int d^3r A_3^0&=&{\rm sgn}(B)\frac{1}{2}\sum_{\mu\ne{\rm val}}
\frac{\langle{\rm val}|
\mbox{\boldmath $\tau$}\cdot\mbox{\boldmath $\Omega$}
|\mu\rangle}{\epsilon_{\rm val}-\epsilon_\mu}
\langle\mu|\sigma_3|{\rm val}\rangle\ +\ {\rm c.\ c.}
\nonumber \\
&{\buildrel{(\ref{defspin})}\over\longrightarrow}&
{\rm sgn}(B)\frac{J_i}{2\alpha^2}\sum_{\mu\ne{\rm val}}
\frac{\langle{\rm val}|\tau_i|\mu\rangle}
{\epsilon_{\rm val}-\epsilon_\mu}
\langle\mu|\sigma_3|{\rm val}\rangle\ +\ {\rm h.\ c.} \ ,
\label{pspin}
\end{eqnarray}
is invariant under $F\rightarrow-F$. This result indicates that the
induced part of the quark current behaves properly under $G$--parity
if there happens to be no ordering ambiguity and the leading term in
the $1/N_C$ expansion vanishes.

\bigskip

\stepcounter{chapter}
\leftline{\large\it 5. Summary and Conclusions}
\bigskip

We have formulated the particle conjugation operation, and especially
its convenient realization as $G$--parity, in the framework of chiral
soliton models. We specifically treated the usual Skyrme model, the
Skyrme model with vector mesons and, as the simplest example
containing quark fields, the chiral quark model. The prescription for
$G$ conjugation in the solitonic sector is simply to reverse the sign of
the profile function $F(r)\rightarrow-F(r)$ while leaving the collective
coordinates $A(t)$ unchanged.

It was found that requiring the correct particle conjugation
properties for currents could resolve operator ordering ambiguities
which greatly affect the results of a number of interesting calculations.
These ambiguities occur in the transition from the classical to the
quantized formulation in the space of the collective coordinates.

In the basic Skyrme model we noted that an hermitean ordering of
operators in the collective space was required to guarantee the
vanishing of a $G$--parity violating axial current form factor.
The same ordering also guaranteed the vanishing of the $G$--parity
violating matrix element of an axial singlet current of interest in
connection with the ``proton spin puzzle".

In the Skyrme model with vector mesons we investigated possible
corrections down by $1/N_C$ to $g_A$ and $\mu_V$. These turned out to
vanish to insure $G$--parity conservation. Generally speaking, our
studies indicate that objects which vanish classically due to
$G$--parity should also vanish in the quantized formulation.

In the chiral quark model the situation turned out to be somewhat
more involved because the quark states are not eigenstates of the
particle conjugation. Having set up the formulation for the
anti--nucleon we were able to show that the next--to leading order
$1/N_C$ corrections to $g_A$ and $\mu_V$, which are non--zero for the
ordering suggested by the {\it ansatz} for the cranked quark fields
(\ref{delval}), behave oppositely to their leading order counterparts
under $F\rightarrow-F$. From this we concluded that the operator
ordering in the space of the collective coordinates should be arranged
so that these corrections vanish. It should be stressed that this
conclusion follows immediately after (\ref{delcur}) without reference to
the quark wavefunction phase convention. This is because the identities
$\alpha_i(\sigma_i){\cal J}=\mp{\cal J}\alpha_i(\sigma_i)$ and
$[{\cal J},\mbox{\boldmath $\tau$}]=0$ immediately show that the leading
and next--to leading order piece of the quark currents in the $1/N_C$
expansion transform oppositely. Since the leading order has the correct
$G$--parity behavior this type of next--to leading order corrections
should actually be absent. This can be accomplished by adopting the
hermitean ordering prescription (\ref{final}) when going from the
classical to the quantum level.

The results we have obtained in the chiral quark model are based on
an ordinary perturbation expansion which is essentially equivalent
to the cranking procedure of ref. \cite{co86}. Using a somewhat
different expansion scheme for the Dirac operator including the
Coriolis term the authors of refs. \cite{bl93,ch94} obtained a
somewhat different result\footnote{We are grateful to C. Christov for
pointing out this difference to us.} for the $1/N_C$ correction to the
axial charge
\begin{eqnarray}
-{\rm sgn}(B)\frac{iN_C}{3\alpha^2}
\sum_{\mu\ne{\rm val}}{\rm sgn}(\epsilon_\mu)
\frac{\langle{\rm val}|\tau_1|\mu\rangle
\langle\mu|\sigma_3\tau_2|{\rm val}\rangle}
{\epsilon_{\rm val}-\epsilon_\mu}.
\label{ga1bo}
\end{eqnarray}
Although this result deviates from the correction in (\ref{gachqu})
only slightly numerically it is obvious from our discussion
that (\ref{ga1bo}) transforms properly under the particle conjugation
due to the additional factor ${\rm sgn}(\epsilon_\mu)$\cite{ch94a}.
It should be remarked that the treatment of refs. \cite{bl93,ch94}
is not itself free of ordering ambiguities, which for example manifest
themselves in the violation of the $PCAC$ relation. Furthermore the
expression (\ref{ga1bo}) cannot be obtained from the ordinary
perturbation expansion or the cranking procedure of ref. \cite{co86}.
It seems to us that a number of open questions remain as to the
validity of (\ref{ga1bo}). The discussion of these goes beyond the
scope of this paper nevertheless our considerations on the particle
conjugation symmetry favor the results of refs. \cite{bl93,ch94} over
those of refs. \cite{wa93,ho93}. The latter are essentially identical
to (\ref{gachqu}). Stated otherwise, the considerations concerning
the particle conjugation symmetry rule out the results of refs.
\cite{wa93,ho93} while no definite statement can be made on those of
refs. \cite{bl93,ch94}. As the models \cite{bl93,ch94} contain
non--local effects they are somewhat different from the chiral quark
model considered here.

The considerations presented in this paper seem to exclude the emergence
of $1/N_C$ corrections to $g_A$ and $\mu_V$ in the context of the
collective quantization of the static soliton configuration when
performing a ``classical" expansion in the angular velocity
$\mbox{\boldmath $\Omega$}$ in local models. This suggests that the
incorporation of pion fluctuations off the soliton is important in
order to obtain non--vanishing corrections. Such a treatment seems
quite appealing since it has been shown to solve problems \cite{wa91}
related to the fact that the current algebra relations cannot be
correctly described within the collective quantization of the static
soliton \cite{ka88}.

A preliminary version of this paper has been submitted elsewhere
\cite{sch94}.

\vskip1cm
\leftline{\it Acknowledgements}
We would like to thank R. Alkofer for helpful discussions.
One of us (HW) acknowledges support by a Habilitanden--scholarship
of the Deutsche Forschungsgemeinschaft (DFG).
This work was supported in part by the US DOE contract number
DE-FG-02-85ER 40231.
\vskip1cm

\appendix
\stepcounter{chapter}
\leftline{\large\it Appendix A: Explicit Solution to
the Chiral Quark Model}
\bigskip

In this appendix we present the explicit form of the eigenfunctions
of the Hamiltonians $h(F)$ and $h(-F)$ for the nucleon and
anti--nucleon, respectively. The Hamiltonian $h(F)$ is given
in (\ref{hstat}). Also the expressions for the axial
charge, $g_A$, the isovector part of the magnetic moment, $\mu_V$,
as well the corresponding ``would--be" $1/N_C$ corrections
$g_A^{(1)}$ and $\mu_V^{(1)}$ are made explicit.

Although only states with grand spin $G=0$ and $G=1$ are relevant
in the framework of the chiral quark model, it is helpful to define
general eigenstates of the grand spin operator
$\mbox{\boldmath $G$}$ (\ref{gspin}):
\begin{eqnarray}
|ljGM\rangle = \sum_{j_3s_3}C^{GM}_{jj_3,\frac{1}{2}s_3}
\sum_{mi_3}C^{jj_3}_{lm,\frac{1}{2}i_3}|lm\rangle
|\frac{1}{2}s_3\rangle_J |\frac{1}{2}i_3\rangle_I.
\label{gstate}
\end{eqnarray}
Here $C^{JM}_{jj_3,j^\prime j_3^\prime}$ denote the SU(2) Clebsch--Gordan
coefficients. The subscripts $J$ and $I$ indicate spinors in
coordinate-- and iso--space, respectively.

\smallskip

\leftline{\it 1. Treatment of the Nucleon}

\smallskip

The quark state which is bound in the chiral background field
$F(r)$ resides in the $G=0$ channel and possesses positive parity
when the boundary condition $F(r=0)=-\pi$ is chosen. In terms of
the grand spin states (\ref{gstate}) this valence quark state
is parametrized as
\begin{eqnarray}
|{\rm val},+\rangle=
\pmatrix{ig^+_0(r)|0\frac{1}{2}00\rangle \cr \cr
f^+_0(r)|1\frac{1}{2}00\rangle \cr}.
\label{val+}
\end{eqnarray}
The radial functions $g_0(r)$ and $f_0(r)$ are obtained by diagonalizing
the Dirac Hamiltonian (\ref{hstat}) and identifying the state with
the energy eigenvalue in the interval $-m\le\epsilon^+_{\rm val}\le m$
as the valence quark. Technically the diagonalization is achieved
by constraining the system to a spherical box of finite radius $D$
and imposing certain boundary conditions at $r=D$. Here we demand that
the upper components of the Dirac spinor (\ref{val+}) vanish at $r=D$.
For more details on the numerical treatment we refer to ref.\cite{re94}.

The matrix elements relevant for the classical parts of the
axial charge and the iso--vector part of the magnetic moments are
easily obtained to be
\begin{eqnarray}
\langle{\rm val},+|\sigma_3\tau_3|{\rm val},+\rangle&=&
\frac{1}{3}\int dr r^2
\left(g^+_0(r)^2-\frac{1}{3}f^+_0(r)^2\right)
\nonumber \\
\langle{\rm val},+|\left[\mbox{\boldmath $r$}\cdot
\mbox{\boldmath $\alpha$},\sigma_3\right]\tau_b
|{\rm val},+\rangle&=&\frac{8i}{3}\delta_{3b}
\int dr r^3 g^+_0(r)f^+_0(r)
\label{stat+}
\end{eqnarray}

In order to evaluate the perturbated wave--function due to cranking
we need to compute the overlap matrix element
$\langle\mu|\tau_j|{\rm val},+\rangle$. The isospin matrix
$\tau_j$ carries unit grand spin and positive parity. Thus
non--vanishing matrix elements exist only when $|\mu\rangle$
belongs to the $1^-$ channel. These states are parametrized as
\begin{eqnarray}
|\mu,1M,-\rangle=
\pmatrix{ig^-_{1\mu}(r)|2\frac{3}{2}1M\rangle \cr \cr
f^-_{1\mu}(r)|1\frac{3}{2}1M\rangle \cr}+
\pmatrix{ig^-_{2\mu}(r)|0\frac{1}{2}1M\rangle \cr \cr
-f^-_{2\mu}(r)|1\frac{1}{2}1M\rangle \cr}
\label{mu-}
\end{eqnarray}
The corresponding eigenvalues of $h(F)$ are denoted by
$\epsilon_\mu^-$. Furthermore $M=-1,0,1$ refers to the projection
of the grand spin $G=1$. Again we refer the interested reader to
ref.\cite{re94} on details on the numerical evaluation of the radial
functions $g^+_{1\mu},..,f^+_{2\mu}$. The boundary conditions on
these radial functions are such that the upper component of the
spinor (\ref{mu-}) vanishes at $r=D$ in order to avoid finite size
isospin violations \cite{we92}.

It is then straightforward to compute the matrix elements relevant
for the axial charge and iso--vector part of the magnetic moments,
which are discussed in section 4:
\begin{eqnarray}
\langle\mu,1M,-|\tau_j|{\rm val|,+}\rangle&=&
\frac{1}{2}S_\mu^+\left[(\tau_j^{11}-\tau_j^{22})\delta_{0M}
-\sqrt{2}\tau_j^{12}\delta_{1M}+
\sqrt{2}\tau_j^{21}\delta_{-1M}\right]
\nonumber \\
\langle{\rm val|,+}|\sigma_3\tau_b|\mu,1M,-\rangle&=&
\frac{1}{3\sqrt{2}}L_\mu^+\left[\tau_b^{12}\delta_{-1M}+
\tau_b^{21}\delta_{1M}\right]
\label{crank+} \\
\langle{\rm val},+|\left[\mbox{\boldmath $r$}\cdot
\mbox{\boldmath $\alpha$},\sigma_3\right]\tau_b|\mu,1M,-\rangle&=&
\frac{-i}{3\sqrt{2}}R_\mu^+\left[\tau_b^{12}\delta_{-1M}+
\tau_b^{21}\delta_{1M}\right].
\nonumber
\end{eqnarray}
The coefficients $S_\mu^+,L_\mu^+$ and $R_\mu^+$ are given as
integrals over the radial functions defined in eqns (\ref{stat+})
and (\ref{crank+}):
\begin{eqnarray}
S_\mu^+&=&-\int dr r^2
\left[g_0^+(r)g_{2\mu}^-(r)+f_0^+(r)f_{2\mu}^-(r)\right]
\nonumber \\
L_\mu^+&=&\int dr r^2
\left[3g_0^+(r)g_{2\mu}^-(r)
-f_0^+(r)\left(f_{2\mu}^-(r)+\sqrt{2}f^-_{1\mu}(r)\right)\right]
\label{radint+} \\
R_\mu^+&=&\int dr r^3 \left[
g_0^+(r)\left(4f_{2\mu}^-(r)+\sqrt{2}f_{1\mu}(r)\right)+
f_0^+(r)\left(4g_{2\mu}^-(r)-\sqrt{2}g_{1\mu}(r)\right)\right].
\nonumber
\end{eqnarray}
The summation over the grand spin projection yields
\begin{eqnarray}
\sum_M\langle{\rm val,+}|\sigma_3\tau_b|\mu,1M,-\rangle
\langle\mu,1M,-|\tau_j|{\rm val,+}\rangle&=&
\frac{i}{3}S_\mu^+L_\mu^+\epsilon_{3bj}
\label{sproj+} \\
\sum_M\langle{\rm val},+|\left[\mbox{\boldmath $r$}\cdot
\mbox{\boldmath $\alpha$},\sigma_3\right]\tau_b|\mu,1M,-\rangle
\langle\mu,1M,-|\tau_j|{\rm val,+}\rangle
&=&\frac{1}{3}S_\mu^+R_\mu^+\epsilon_{3bj}
\nonumber
\end{eqnarray}

\smallskip

\leftline{\it 2. Treatment of the Anti--Nucleon}

\smallskip

For the boundary condition $F(r=0)=\pi$ the parity of the bound
valence quark gets reversed. Thus this state is member of the
$0^-$ channel:
\begin{eqnarray}
|{\rm val},-\rangle=
\pmatrix{ig^-_0(r)|1\frac{1}{2}00\rangle \cr \cr
f^-_0(r)|0\frac{1}{2}00\rangle \cr}.
\label{val-}
\end{eqnarray}
The corresponding eigenvalue of the Dirac Hamiltonian (\ref{hstat})
is labeled $\epsilon_{\rm val}^-$. As the transformation ${\cal J}$
in eqn (\ref{hrevf}) exchanges upper and lower components of a Dirac
spinor, we now demand the lower components of (\ref{val-}) to vanish
at $r=D$. This makes possible the comparison with (\ref{val+}) even
for finite $D$.

The classical parts of $g_A$ and $\mu_V$ are given by
\begin{eqnarray}
\langle{\rm val},-|\sigma_3\tau_3|{\rm val},-\rangle&=&
-\frac{1}{3}\int dr r^2
\left(f^-_0(r)^2-\frac{1}{3}g^-_0(r)^2\right)
\nonumber \\
\langle{\rm val},-|\left[\mbox{\boldmath $r$}\cdot
\mbox{\boldmath $\alpha$},\sigma_3\right]\tau_b
|{\rm val},-\rangle&=&\frac{8i}{3}\delta_{3b}
\int dr r^3 f^-_0(r)g^-_0(r)
\label{stat-}
\end{eqnarray}
Under the cranking perturbation the valence quark state (\ref{val-})
couples to the states in the $1^+$ channel
\begin{eqnarray}
|\mu,1M,+\rangle=
\pmatrix{ig^+_{1\mu}(r)|1\frac{3}{2}1M\rangle \cr \cr
-f^+_{1\mu}(r)|2\frac{3}{2}1M\rangle \cr}+
\pmatrix{ig^+_{2\mu}(r)|1\frac{1}{2}1M\rangle \cr \cr
f^+_{2\mu}(r)|0\frac{1}{2}1M\rangle \cr}.
\label{mu+}
\end{eqnarray}
Their eigenvalues are denoted by $\epsilon_{\mu}^+$.
Again the lower components are constrained to be zero at $r=D$.
The isospin structure of the matrix elements (\ref{crank+})
remains unaltered when we transfer them to the anti--nucleon.
However, the coefficients $S_\mu^+,L_\mu^+$ and $R_\mu^+$ are
replaced by
\begin{eqnarray}
S_\mu^-&=&-\int dr r^2
\left[g_0^-(r)g_{2\mu}^+(r)+f_0^-(r)f_{2\mu}^+(r)\right]
\nonumber \\
L_\mu^-&=&\int dr r^2
\left[3f_0^-(r)f_{2\mu}^+(r)
-g_0^-(r)\left(g_{2\mu}^+(r)-\sqrt{2}g^+_{1\mu}(r)\right)\right]
\label{radint-} \\
R_\mu^-&=&\int dr r^3 \left[
g_0^-(r)\left(4f_{2\mu}^+(r)+\sqrt{2}f^+_{1\mu}(r)\right)+
f_0^-(r)\left(4g_{2\mu}^+(r)-\sqrt{2}g^+_{1\mu}(r)\right)\right].
\nonumber
\end{eqnarray}
When the summation over the grand spin projection is performed these
integrals have to be substituted for their analogous expressions in
eqn (\ref{sproj+}).

Using the explicit representations
(\ref{val+}, \ref{mu-}, \ref{val-}) and (\ref{mu+}) the
transformation (\ref{statrev}) corresponds to
\begin{eqnarray}
g_0^+(r)\rightarrow -f_0^-(r),\qquad
f_0^+(r)\rightarrow g_0^-(r),
\nonumber \\
g_1^-(r)\rightarrow f_1^+(r),\qquad
f_1^-(r)\rightarrow -g_1^+(r),
\label{rfctrev} \\
g_2^-(r)\rightarrow -f_2^+(r),\qquad
f_2^-(r)\rightarrow g_2^+(r).
\nonumber
\end{eqnarray}

Numerically we have verified that the self--consistent solution
reverses its sign ($F\rightarrow-F$) when we alter the boundary
condition from $F(0)=-\pi$ to $F(0)=\pi$. Furthermore the
computations confirm the transformations
$\epsilon_{\rm val}^+\rightarrow-\epsilon_{\rm val}^-$ and
$\epsilon_\mu^-\rightarrow-\epsilon_\mu^+$ for the eigenvalues of
$h(F)$. The transformation properties for the wave--functions
(\ref{rfctrev}) are regained up to an overall, irrelevant phase.
This phase may, of course, be different in the grand spin zero and
one channels.

\bigskip

\stepcounter{chapter}
\leftline{\large\it Appendix B: Completeness Argument}

\bigskip

Using the results from appendix A the transformation relations
(\ref{leading}) and (\ref{next}) are easily verified.
In terms of the integrals over the wave--functions these
transformations are expressed as
\begin{eqnarray}
S_\mu^+L_\mu^+\rightarrow S_\mu^-L_\mu^-
\quad {\rm and}\quad
S_\mu^+R_\mu^+\rightarrow -S_\mu^-R_\mu^-.
\qquad({\rm no\ sum\ over}\ \mu)
\label{inttrans}
\end{eqnarray}
The validity of the relations (\ref{leading}) and (\ref{next})
is also nicely confirmed by making use of the completeness property
of the eigenstates of the Dirac Hamiltonian. From that one can deduce
\begin{eqnarray}
\langle{\rm val},\pm|\sigma_3\tau_3|{\rm val},\pm\rangle
=\frac{1}{3}\sum_\mu S_\mu^\pm L_\mu^\pm
\quad {\rm and}\quad
\langle{\rm val},\pm|\left[\mbox{\boldmath $r$}\cdot
\mbox{\boldmath $\alpha$},\sigma_3\right]\tau_3
|{\rm val},\pm\rangle=
\frac{i}{3}\sum_\mu S_\mu^\pm R_\mu^\pm.
\label{complete}
\end{eqnarray}
This indicates once again that the matrix elements involved
in the subleading order of the $1/N_C$ expansion transform
identically to those of the leading order when the sign of the
chiral angle is reversed. We should mention that we have
confirmed the identities (\ref{complete}) numerically for both
boundary conditions $F(0)=\pm\pi$.

\vfil\eject

\end{document}